\documentclass[aps,prb,superscriptaddress,twocolumn,floatfix,showpacs,citeautoscript,longbibliography,hyperlinks]{revtex4-2}
\usepackage{amsmath}
\usepackage{amssymb}
\usepackage{amsfonts}
\usepackage{dsfont}
\usepackage{color}
\usepackage[autostyle]{csquotes}
\usepackage[colorlinks=true,citecolor=blue]{hyperref}
\usepackage{graphicx,graphics}

\usepackage{xcolor}
\newcommand{\revision}[1]{{{#1}}}

\begin{document}
\title{Controlled emission time	statistics of a dynamic single-electron transistor}

\author{Fredrik~Brange}
\affiliation{Department of Applied Physics, Aalto University, 00076 Aalto, Finland}
\author{Adrian~Schmidt}
\affiliation{Institut f\"ur Festk\"orperphysik, Leibniz Universit\"at Hannover, Hannover, Germany}
\author{Johannes~C.~Bayer}
\affiliation{Institut f\"ur Festk\"orperphysik, Leibniz Universit\"at Hannover, Hannover, Germany}
\author{Timo~Wagner}
\affiliation{Institut f\"ur Festk\"orperphysik, Leibniz Universit\"at Hannover, Hannover, Germany}
\author{Christian~Flindt}
\affiliation{Department of Applied Physics, Aalto University, 00076 Aalto, Finland}
\author{Rolf~J.~Haug}
\affiliation{Institut f\"ur Festk\"orperphysik, Leibniz Universit\"at Hannover, Hannover, Germany}

\date{\today}

\begin{abstract}
Quantum technologies involving qubit measurements based on electronic interferometers rely critically on accurate
single-particle emission. However, achieving precisely timed operations requires exquisite control of the single-particle
sources in the time domain. Here, we demonstrate accurate control of the emission time statistics of a
dynamic single-electron transistor by measuring the waiting times between emitted electrons. By ramping up the modulation frequency, we controllably drive the system through a crossover from adiabatic to nonadiabatic dynamics, which we visualize by measuring the temporal fluctuations at the single-electron level and explain using detailed theory. Our work paves the way for future technologies based on the ability to control, transmit, and detect single quanta of charge or heat in the form of electrons, photons, or phonons.
\end{abstract}

\maketitle

\textbf{Introduction.} 
Understanding the interplay between driving frequency and response time is of critical importance for quantum technologies  that require carefully timed operations such as in qubit measurements via interferometric setups \cite{Splettstoesser2017}. In such applications, quantum interference is only observed if individual charges emitted from separate single-electron sources arrive simultaneously at an electronic beamsplitter \cite{Feve2007,Bocquillon2013a,Bocquillon2013b,Dubois2013,Jullien2014,Ubbelohde2014}. Similar requirements appear for metrological current standards where a precisely defined current is obtained only if exactly one electron is emitted per period  \cite{Blumenthal2007,Pekola2013,Kaestner2015}. Optimal control of the single-electron sources is thus an important prerequisite for quantum technologies operating with fixed clock cycles. For the analysis of dynamic processes, measurements of the waiting time \cite{vanKampen2007,PhysRevA.39.1200} between emitted particles have been suggested  \cite{Brandes2008,Albert2011,Dasenbrook2014,Potanina2017,Burset2019,Rudge2019}. Measuring the waiting time distribution, however, is challenging, since it requires nearly perfect detectors and high statistical accuracy. Still, experiments~\cite{Delteil2014,PhysRevApplied.8.034019,PhysRevResearch.1.033163} are motivated by the prospects of analyzing dynamic processes in the time domain \cite{Brandes2008,Albert2011,Dasenbrook2014,Potanina2017,Burset2019,Rudge2019}.

In this work, we demonstrate accurate experimental control of the temporal statistics of electrons emitted from a periodically driven single-electron transistor. By modulating the applied gate-voltage periodically in time, we modify the rates at which electrons tunnel in and out of the single-electron transistor, and we are thereby able to reliably control the resulting emission-time statistics. In order to analyze the temporal fluctuations, we use a sensitive single-electron detector to precisely meausure the full distribution of waiting times between emitted electrons, allowing us to visualize a controlled crossover from adiabatic to nonadiabatic dynamics as we ramp up the external driving frequency.

\begin{figure*}
	\includegraphics[width=\textwidth]{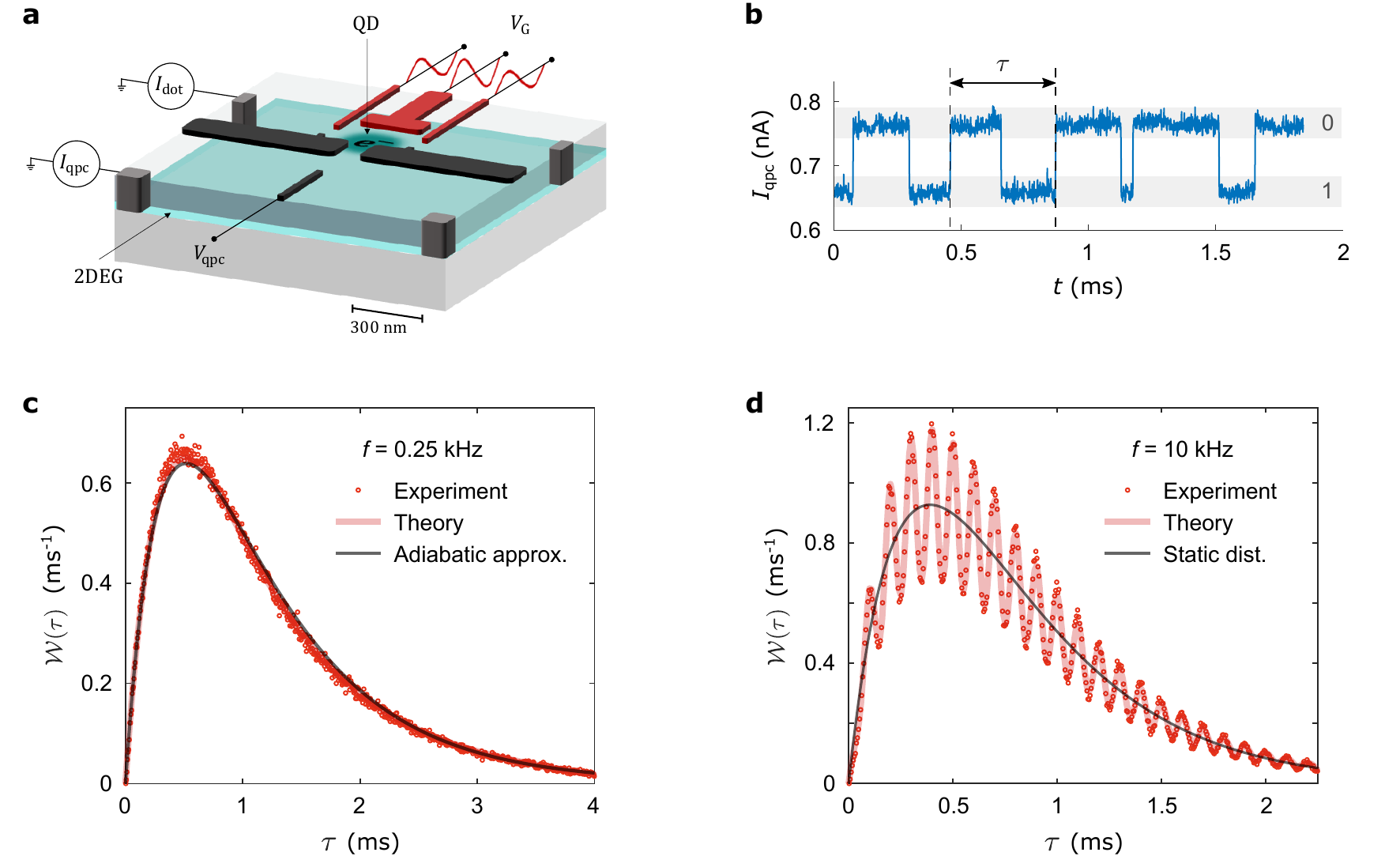}
	\caption{\textbf{Dynamic single-electron transistor and waiting time distributions.} \textbf{a,} Schematic of the gate-defined quantum dot in a two-dimensional electron gas (2DEG) tunnel-coupled to two external electrodes. A harmonic voltage is applied to the gate electrodes in red, which modulates the tunneling of single electrons between the quantum dot and the leads. A capacitively coupled quantum point contact  is used to monitor the charge state of the quantum dot. \textbf{b,} Time-trace of the current in the quantum point contact, which switches between two distinct levels corresponding to having 0 or 1 electron on the quantum dot. The waiting time between electrons tunneling out of the quantum dot is denoted by $\tau$. \textbf{c,} Distribution of waiting times measured at the driving frequency $f= 0.25$ kHz together with exact calculations and the adiabatic approximation~(\ref{Adiabatic approximation}). \textbf{d,} Distribution measured at the driving frequency $f=10$ kHz together with exact calculations and the static expression~(\ref{Static WTD}) with period-averaged rates inserted.
		\label{Fig1}}
\end{figure*}

Figure~\ref{Fig1}a shows the dynamic single-electron transistor consisting of a nano-scale quantum dot coupled to external electrodes defined by electrostatic gating of a two-dimensional electron \revision{gas.} In Ref.~\cite{Wagner2019}, the device was used to observe a stochastic resonance in a periodically driven quantum dot. One panel shows a distribution of residence times, which quantify how long the quantum dot is occupied. The waiting time, by contrast, measures the time span between two subsequent electron emissions, which is useful to characterize the regularity of dynamic single-electron emitters as discussed here. The system is operated in the Coulomb blockade regime, where the quantum dot can be occupied by only zero or one electron at a time. In addition to a small voltage, $V = 1$~mV, we apply a harmonic drive to the gate electrodes, $V_G(t) = \Delta V \sin(2\pi f t)$, with amplitude $\Delta V = 10$~mV and adjustable frequency $f$ in the kilohertz range. The gate voltage modulates the tunneling of electrons in and out of the quantum dot with rates that to a good approximation depend exponentially on the \revision{external driving}~\cite{MacLean2007} as $\Gamma_i(t) = \Gamma_i \exp[\alpha_i \sin(2\pi ft)]$ and $\Gamma_o(t) = \Gamma_o\exp[-\alpha_o \sin(2\pi ft)]$, where $\alpha_o=0.81$, $\alpha_i=0.64$, $\Gamma_o=1.8$~kHz, and $\Gamma_i\approx 2.1$~kHz is weakly frequency-dependent (see Materials and Methods). To detect the individual  tunneling events, we measure a separate electrical current that runs through a capacitively coupled quantum point contact, whose conductance depends sensitively on the occupation of the quantum dot.

Figure \ref{Fig1}b shows a typical time-trace of the current in the quantum point contact, illustrating how it switches between two distinct levels in real-time, signaling that single electrons tunnel in and out of the quantum dot. To analyze the response of the system to the external drive, we measure the waiting times $\tau$ between individual electrons tunneling out of the quantum dot \cite{Brandes2008,Albert2011,Dasenbrook2014,Potanina2017,Burset2019,Rudge2019}.

\textbf{Results.} Figure \ref{Fig1}c shows the distribution of waiting times, collected from about $10^6$ detected tunneling events during a measurement time of approximately 10 minutes and a driving frequency of $f= 0.25$~kHz. The waiting time distribution is suppressed to zero at short times, since the quantum dot cannot by doubly occupied, and the strong Coulomb interactions thereby prevent two electrons from leaving the quantum dot simultaneously. At later times, the quantum dot can be refilled, and the suppression is gradually lifted with the distribution peaking at around $\tau\simeq 0.6$~ms, before it vanishes at much longer times. This behavior is very different from a Poisson process, such as the decay of radioactive nuclei at rate $\Gamma$, for which the distribution of waiting times is exponential, $\mathcal{W}(\tau) = \Gamma e^{-\Gamma \tau}$. Indeed, with constant rates, $\Gamma_i(t) = \Gamma_i$ and $\Gamma_o(t) = \Gamma_o$, the distribution would read~\cite{Brandes2008}
\begin{equation}
\mathcal{W}_s(\tau,\Gamma_{i},\Gamma_{o}) = \frac{\Gamma_{i}\Gamma_{o}}{\Gamma_{i}-\Gamma_{o}}\left( e^{-\Gamma_{o} \tau}- e^{-\Gamma_{i} \tau}\right)
\label{Static WTD}
\end{equation}
with $\mathcal{W}_s(\tau,\Gamma,\Gamma) =\Gamma^2 \tau e^{-\Gamma \tau}$ for equal tunneling rates. In the experiment, the rates are time-dependent, but in Fig.~\ref{Fig1}c the driving frequency is much lower than the typical tunneling rates, $f\ll\Gamma_i,\Gamma_o$, and we expect that the system will adiabatically follow the external modulations. In that case, the waiting time distribution should be given by a period-average over the static distribution~(\ref{Static WTD}) with the time-dependent rates $\Gamma_i(t)$ and $\Gamma_o(t)$ inserted~\cite{Potanina2017}
\begin{equation}
\mathcal{W}(\tau) = \int_0^T \frac{dt}{T} \mathcal{W}_s(\tau,\Gamma_i(t),\Gamma_o(t)),
\label{Adiabatic approximation}
\end{equation}
where $T=1/f$ is the period of the drive. This adiabatic approximation agrees very well with the measurements, and it demonstrates that the system is in sync with the external drive, and the dynamic response is adiabatic. 

\begin{figure*}
	\includegraphics[width=0.99\textwidth]{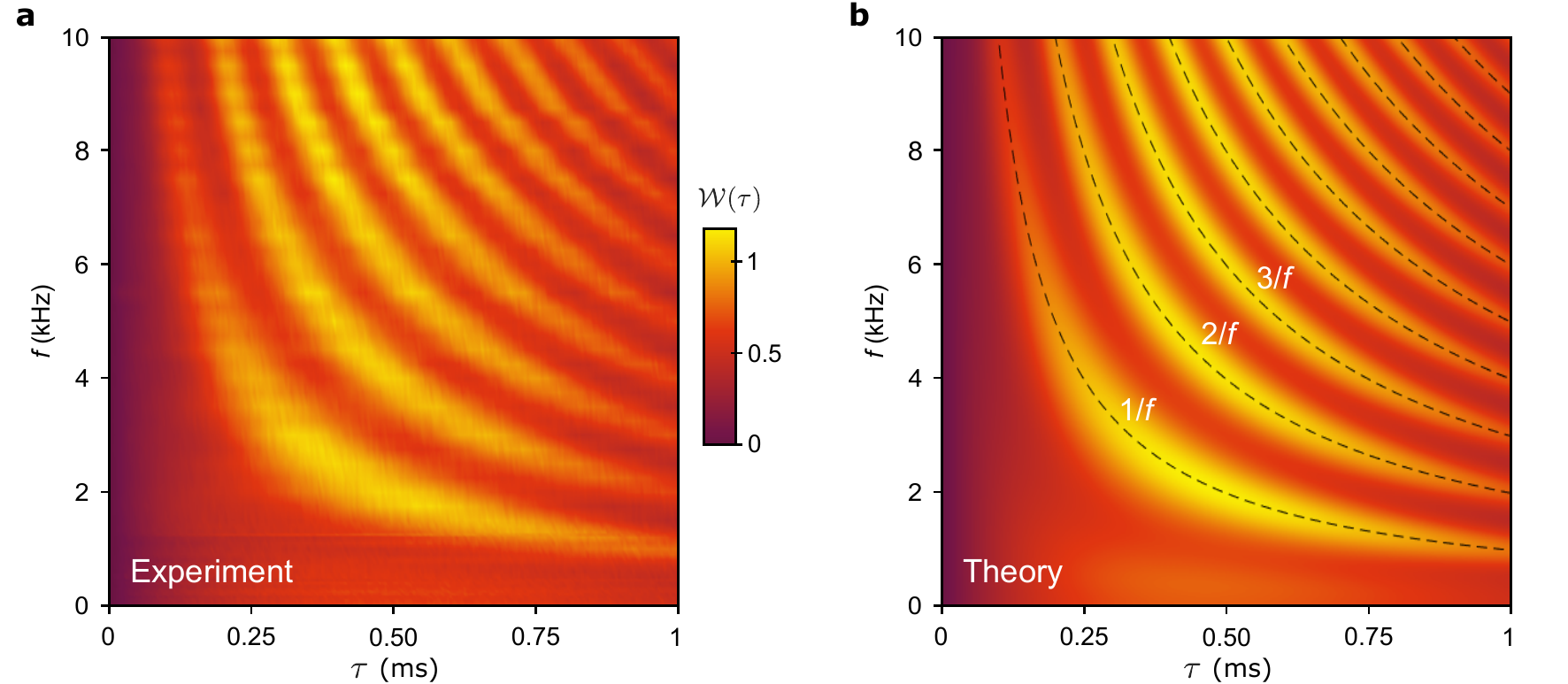}
	\caption{\textbf{Distributions of waiting times as functions of the driving frequency.} \textbf{a,} Measured waiting time distributions with varying driving frequency $f$. \textbf{b,} Calculations of  waiting time distributions with the parameters $\alpha_o = 0.81$, $\alpha_i = 0.64$, $\Gamma_o = 1.8$~kHz, and $\Gamma_i= 2.3$~kHz. The dashed lines indicate the peaks in the distributions at $\tau_n=n/f$ according to Eq.~(\ref{Integer times}).
		\label{Fig2}}
\end{figure*}

In Fig.~\ref{Fig1}d, we have ramped \revision{up} the driving frequency to $f = 10$ kHz, and a completely different picture now emerges. The driving frequency is much faster than the tunneling rates, and the system can no longer follow the fast high-frequency modulations. One might expect that the waiting time distribution would be given by the static result~(\ref{Static WTD}) with period-averaged rates, $\overline \Gamma_{i,o} = \int_0^T dt \Gamma_{i,o}(t)/T$, inserted. Indeed, the overall curve follows this result as shown with a black line in Fig.~\ref{Fig1}d. However, the distribution exhibits an oscillatory pattern on top of the static result, and a detailed theoretical analysis of the high-frequency regime yields the expression~\footnote{See supplementary information.}
\begin{equation}
\mathcal{W}(\tau) = \mathcal{W}_s(\tau,\overline{\Gamma}_i,\overline{\Gamma}_o)\left(1+\frac{\alpha_o^2}{2} \cos(2\pi f \tau)\right),
\label{eq:hf_wtd}
\end{equation}
which is valid up to second order in $\alpha_i$ and $\alpha_o$. Although higher-order corrections are needed to fully capture the experimental results, this expression explains the oscillations in the waiting time distribution with peaks occuring at multiples of the period as seen in the figure,
\begin{equation}
\tau_n = n T=n/f, \quad n = 1,2,3,...,
\label{Integer times}
\end{equation}
indicating that the driving is now highly nonadiabatic. In Fig.~\ref{Fig1}, we also show exact calculations~\cite{Note1} (with no adjustable parameters) that are in excellent agreement with the measurements and thus support our interpretations.

\begin{figure*}
	\includegraphics[width=\textwidth]{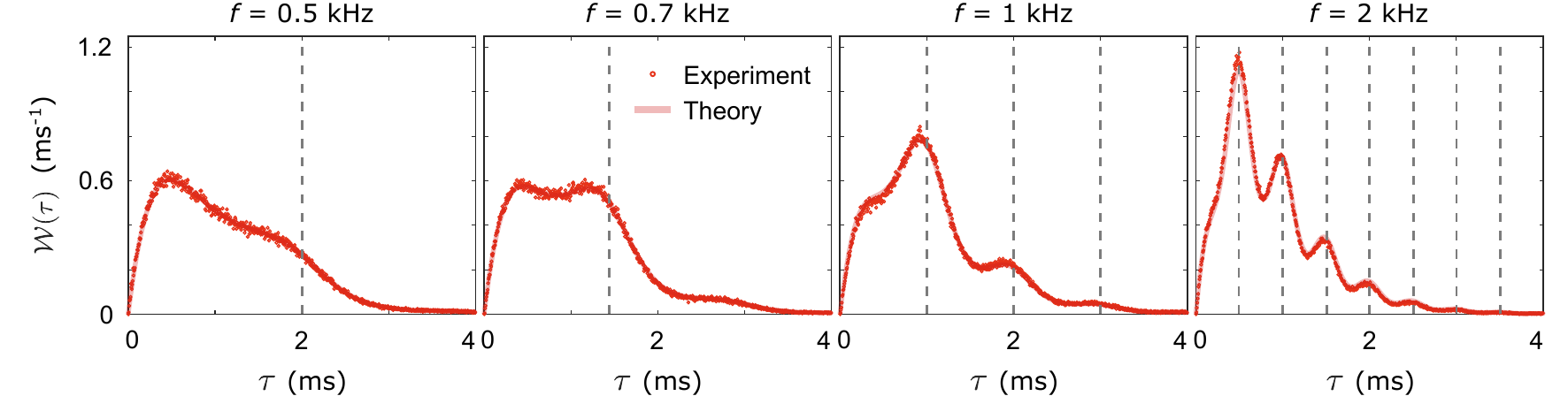}
	\caption{\textbf{Adiabatic-to-nonadiabatic crossover.} Distributions of electron waiting times for four driving frequencies in the crossover region, $f=0.5$ kHz (left), $0.7$ kHz, $1$ kHz, and  $2$ kHz (right). Vertical dashed lines indicate multiplies of the period.
		\label{Fig3}}
\end{figure*}

To characterize the crossover from adiabatic to nonadiabatic dynamics and thereby demonstrate  full control of the emission-time statistics, waiting time distributions across the whole range of driving frequencies are displayed in Fig.~\ref{Fig2}. The left panel shows experimental results for a wide range of driving frequencies, while the right panel contains the corresponding calculations of the waiting time distributions. The figure clearly illustrates how the oscillatory pattern in the waiting time distributions builds up with increasing driving frequency, and it corroborates the physical picture that peaks should appear at multiples of the driving period  according to Eq.~(\ref{Integer times}). From our theoretical analysis, we anticipate that the crossover from adiabatic to nonadiabatic dynamics will take place for driving frequencies that are on the order of the tunneling rates; a regime, where a stochastic resonance also occurs \cite{Wagner2019}. To explore the crossover in detail, Fig.~\ref{Fig3} diplays distributions in this frequency range.

The leftmost panel of Fig.~\ref{Fig3} shows the waiting time distribution for $f=0.5$~kHz. Here, the distribution is still dominated by the adiabatic peak at short waiting times, however, a small shoulder developing at the period of the drive provides the first indications of a nonadiabatic response. In the next panel, the frequency has been increased to $f=0.7$~kHz, and a peak is now becoming visible at the period of the drive together with a shoulder at twice the period. In the third panel, we have further increased the frequency to $f=1$ kHz, and the waiting time distribution is now  distinctly dominated by peaks at multiples of the period, signaling that we are reaching the nonadiabatic regime. Finally, in the rightmost panel with $f=2$ kHz, the waiting time distribution is completely governed by the nonadiabatic peak structure, and we no longer see traces of the adiabatic distribution.

\textbf{Discussion.} Our work demonstrates unprecedented control of the emission-time statistics of a dynamic single-electron transistor. By increasing the external modulation frequency, we have carefully driven the system through a crossover from adiabatic to non-adiabatic dynamics, which could be clearly visualized in measurements of the electron waiting time distribution. We have thus established waiting time distributions as an important experimental concept in the time-domain analysis of dynamic single-particle sources, not only for those that emit electrons \cite{Brandes2008,Rudge2019,Albert2011,Dasenbrook2014,Potanina2017,Burset2019} but also for systems involving other discrete quanta such as single photons \cite{Aharonovich2016} or phonons \cite{Cohen2015}. While we have considered tunneling of confined electrons in a low-dimensional structure, future experiments may measure the waiting times between charge pulses propagating in extended electronic wave guides \cite{Feve2007,Bocquillon2013a,Bocquillon2013b,Dubois2013,Jullien2014}. In combination, these efforts are important for future technologies operating with fixed clock cycles such as interferometric devices \cite{Feve2007,Bocquillon2013a,Bocquillon2013b,Dubois2013,Jullien2014,Ubbelohde2014} and metrological current standards \cite{Blumenthal2007,Pekola2013,Kaestner2015}. \\

\noindent {\textbf{Materials and Methods}} 

\noindent {\textbf{Experimental details.}} The device is based on a GaAs/AlGaAs heterostructure with a two-dimensional electron gas (2DEG). The 2DEG is formed 100 nm below the surface of the heterostructure and has a charge density of $2.4 \cdot 10^{11}$ cm$^{-2}$ with a mobility of $5 \cdot 10^5$ cm$^2$V$^{-1}$s$^{-1}$. On the surface of the heterostructure, CrAu gates are formed by e-beam and optical lithography. By applying negative gate voltages, the 2DEG below the gates is depleted and the quantum dot as well as the quantum point contact are formed.

The device was operated in a $^4$He cryostat at 1.5 K, while the signal processing was done outside at room temperature. The driving signal was generated with the help of an Adwin Pro2 real-time system. To amplify the current through the quantum point contact, a low-noise amplifier with 100 kHz bandwidth was used. The detector current $I_{\mathrm{qpc}} (t)$ was monitored with a temporal resolution of $\Delta t_s = 2.5$ $\mu$s. To extract the waiting times, the time-dependent occupation of the quantum dot was determined. To this end, the measured traces of $I_{\mathrm{qpc}}(t)$ were digitized with the high current level indictating that the quantum dot was empty (state 0) and the low current level that it was occupied (state 1). The waiting times,~$\tau$, between single electrons tunneling out of the quantum dot were identified as the time between consecutive transitions from state 1 to state 0. The time-dependent tunneling rates $\Gamma_i(t)$ and $\Gamma_o(t)$ were extracted from the experimental data, and the parameters $\alpha_o$, $\alpha_i$, $\Gamma_o$ and $\Gamma_i$ were subsequently determined. For our calculations, we use $\alpha_i = 0.64$, $\alpha_o = 0.81$ and $\Gamma_o = 1.8$~kHz, while $\Gamma_i$ is weakly frequency-dependent as shown below:

\begin{center}
\begin{tabular}{ | c | c |  c | c | c | c | c |  c | c | c | }
		\hline
		$f$ (kHz) & 0.00 & 0.25 & 0.50 & 0.70 & 1.0 & 2.0 & 4.0 & 8.0 & 10 \\ \hline
		$\Gamma_i$ (kHz) &
		1.8 & 1.9 & 1.7 & 1.7 & 1.7 & 2.2 & 2.3 & 2.6 & 2.9 \\
		\hline
\end{tabular}
\end{center}

\noindent {\textbf{Theory.}} The system can be described by the rate equation
$\frac{d}{dt}\left|p(t)\right\rangle = \mathbf{L}(t)\left| p(t)\right\rangle=[\mathbf{L}_0(t)+\mathbf{J}(t)]\left|p(t)\right\rangle$,
where the vector  $\left|p(t)\right\rangle =(p_{0}(t),p_{1}(t))^{\mathrm{T}}$ contains the probabilities for the quantum dot to be empty or occupied, and we have partitioned the rate matrix $\mathbf{L}(t)$ into
\begin{equation}
\mathbf{L}_0(t)=\left( \begin{array}{cc} -\Gamma_{i}(t) & 0 \\ \Gamma_{i}(t) & -\Gamma_{o}(t)  \end{array} \right) \,\,\mathrm{and}\,\,\,
\mathbf{J}(t)=\left( \begin{array}{cc} 0 & \Gamma_{o}(t) \\ 0 & 0 \end{array} \right),
\nonumber
\end{equation}
where $\Gamma_{i,o}(t)$ are the tunneling rates, and $\mathbf{J}(t)$ describes electrons tunneling out of the quantum dot. The waiting time distribution can be calculated as $\mathcal{W}(\tau)=\langle\tau\rangle \partial^2_\tau\Pi(\tau)$, where $\langle\tau\rangle=-1/[\partial_\tau\Pi(0)]$ is the mean waiting time, and $\Pi(\tau)$ is the idle-time probability that no electrons have tunneled out of the quantum dot during a time span of duration~$\tau$ \cite{Dasenbrook2014,Potanina2017}. For a peridically-driven system, this probability depends not only on the length of the interval, $\tau$, but also on the starting point, $t_0$, and we have to average it over a period of the drive as $\Pi(\tau)=\int_0^Tdt_0 \Pi(\tau,t_0)/T$. The idle-time probability can be expressed as
$\Pi(\tau,t_0)=\langle 1|\hat{T}\{e^{\int_{t_0}^{\tau+t_0}dt \mathbf{L}_0(t)} \}\left|p_s(t_0)\right\rangle$,
where $\hat{T}$ is the time-ordering operator, the periodic state is denoted as $|p_s(t_0)\rangle=|p_s(t_0+T)\rangle$, and we have defined $\langle 1|=(1,1)$ \cite{Potanina2017}. The periodic state is found by solving the eigenproblem, $\hat{T}\{e^{\int_{t_0}^{T+t_0}dt \mathbf{L}(t)} \}\left|p_s(t_0)\right\rangle=|p_s(t_0)\rangle$, for $|p_s(t_0)\rangle$ by using the normalization, $p_0(t_0)+p_1(t_0)=1$, at all times.\\

\noindent {\textbf{Acknowledgements}} \\
The work was financially supported by Deutsche Forschungsgemeinschaft (DFG, German Research Foundation) under Germany's Excellence Strategy -- EXC 2123/1 QuantumFrontiers 390837967,  the State of Lower Saxony, Germany, via Hannover School for Nanotechnology and School for Contacts in Nanosystems, and by the Academy of Finland (projects No.~308515, 312057, 312299, and 331737). F.B. acknowledges support from the European Union's Horizon 2020 research and innovation programme under the Marie Sk\l{}odowska--Curie grant agreement No.~892956. \\

\noindent {\textbf{Author contributions}} \\
T.W.~carried out the experiment and A.S.~processed the measurement data with support from J.C.B. The theory was developed by F.B.~and~C.F. All authors participated in the discussions of the results. The manuscript was prepared by F.B.,~C.F.,~A.S.,~and R.J.H. The research was supervised by C.F.~and R.J.H. \\

\noindent {\textbf{Competing interests}} \\
The authors declare that they have no competing interests.

\noindent {\textbf{Data availability}} \\
All data needed to evaluate the conclusions in the paper are present in the paper and  the supplementary material. Additional data related to this paper may be requested from the authors.

\end{document}


\title{Supplementary material: \\ Controlled emission time statistics of a dynamic single-electron transistor}

\author{Fredrik~Brange}
\affiliation{Department of Applied Physics, Aalto University, 00076 Aalto, Finland}
\author{Adrian~Schmidt}
\affiliation{Institut f\"ur Festk\"orperphysik, Leibniz Universit\"at Hannover, Hannover, Germany}
\author{Johannes~C.~Bayer}
\affiliation{Institut f\"ur Festk\"orperphysik, Leibniz Universit\"at Hannover, Hannover, Germany}
\author{Timo~Wagner}
\affiliation{Institut f\"ur Festk\"orperphysik, Leibniz Universit\"at Hannover, Hannover, Germany}
\author{Christian~Flindt}
\affiliation{Department of Applied Physics, Aalto University, 00076 Aalto, Finland}
\author{Rolf~J.~Haug}
\affiliation{Institut f\"ur Festk\"orperphysik, Leibniz Universit\"at Hannover, Hannover, Germany}

\maketitle

\section{Calculations of waiting time distributions}

\subsection{Numerical calculations}

\noindent To calculate the distribution of waiting times, we follow the Materials and Methods section and Ref.~16 of the main text. We first compute the idle-time probability as
$\Pi(\tau,t_0) = \langle 1|\mathbf{U}_{0}(\tau+t_0,t_0)\left|p_s(t_0)\right\rangle$ with $\mathbf{U}_{0}(t,t_0)=\hat{T}\{e^{\int_{t_0}^{t}dt_1 \mathbf{L}_0(t_1)} \}$, where $\hat T$ is the time-ordering operator and $\langle 1|\equiv(1,1)$. For the periodic state, $|p_s(t_0)\rangle=|p_s(t_0+T)\rangle$, we find
\begin{equation}
p_1(t_0) = \frac{e^{-\int_{t_0}^{t_0+T} dt_1 (\Gamma_i(t_1)+\Gamma_o(t_1))}}{1-e^{-\int_{t_0}^{t_0+T} dt_1 (\Gamma_i(t_1)+\Gamma_o(t_1))}} \int_{t_0}^{t_0+T} dt_1 \Gamma_i(t_1) e^{\int_{t_0}^{t_1} dt_2 [\Gamma_i(t_2)+\Gamma_o(t_2)]}.
\label{p0}
\end{equation}
with the normalization, $p_0(t_0)+p_1(t_0)=1$, at all times. For the non-zero matrix elements of $\mathbf{U}_{0}(t,t_0)$, we have$^{16}$
\begin{equation}
\label{eq:U}
\begin{split}
[\mathbf{U}_{0}(t,t_0)]_{11} &= \exp\left[ -\int_{t_0}^t dt_1 \Gamma_i(t_1) \right], \qquad [\mathbf{U}_{0}(t,t_0)]_{22} = \exp\left[ -\int_{t_0}^t dt_1 \Gamma_o(t_1) \right], \\
[\mathbf{U}_{0}(t,t_0)]_{21} &= \exp\left[ -\int_{t_0}^t dt_1 \Gamma_o(t_1) \right] \int_{t_0}^t dt_1 \Gamma_i(t_1) \exp\left[ \int_{t_0}^{t_1} dt_2 \left[ \Gamma_o(t_2)-\Gamma_i(t_2) \right] \right],
\end{split}
\end{equation}
and thereby obtain
\begin{equation}
\Pi(\tau,t_0) = U_{11}(t_0+\tau,t_0)[1-p_1(t_0)]+U_{21}(t_0+\tau,t_0)[1-p_1(t_0)]+U_{22}(t_0+\tau,t_0)p_1(t_0).
\label{Pi calc}
\end{equation}
By evaluating these expressions numerically, we obtain the waiting time distribution as $\mathcal{W}(\tau)=\langle\tau\rangle \partial^2_\tau\Pi(\tau)$ with $\langle\tau\rangle=-1/[\partial_\tau\Pi(\tau=0)]$ and  $\Pi(\tau)=\int_0^Tdt_0 \Pi(\tau,t_0)/T$.

\subsection{Zero-frequency limit}
\noindent With constant rates, $\Gamma_i(t) =\Gamma_i$ and $\Gamma_o(t) = \Gamma_o$, Eqs.~\eqref{p0} and \eqref{eq:U} are easily solved analytically, yielding
\begin{equation}
p_1^s(t_0,\Gamma_i,\Gamma_o) = \frac{\Gamma_i}{\Gamma_i+\Gamma_o},
\end{equation}
and
\begin{equation}
\begin{split}
[\mathbf{U}_{0}(t,t_0,\Gamma_i)]_{11}^s &= \exp\left[-\Gamma_i(t-t_0) \right],\qquad [\mathbf{U}_{0}(t,t_0,\Gamma_o)]_{22}^s = \exp\left[-\Gamma_o(t-t_0) \right], \\
[\mathbf{U}_{0}(t,t_0,\Gamma_i,\Gamma_o)]_{21}^s &= \frac{{\Gamma}_i}{{\Gamma}_i-{\Gamma}_o}\left(e^{-{\Gamma}_o(t-t_0)}-e^{-{\Gamma}_i(t-t_0)} \right).
\end{split}
\end{equation}
We then find
\begin{equation}
\Pi_s(t,\Gamma_i,\Gamma_o) = \frac{e^{-\Gamma_o t}\Gamma_i^2-e^{-\Gamma_i t}\Gamma_o^2}{ \Gamma_i^2- \Gamma_o^2},
\end{equation}
and
\begin{equation}
\mathcal{W}_s(\tau,\Gamma_i,\Gamma_o) = \frac{\Gamma_i\Gamma_o}{\Gamma_i-\Gamma_o} \left( e^{-\Gamma_o\tau}-e^{-\Gamma_i\tau} \right),
\end{equation}
which is Eq.~(1) of the main text.

\subsection{High-frequency limit}
\noindent For $\alpha_i, \alpha_o \ll1$, we may treat the driving as a perturbation and expand all quantities to second order in the driving amplitudes (the first-order contribution to the waiting time distribution eventually vanishes). We start by expanding the tunneling rates as
\begin{equation}
\Gamma_i(t) = \Gamma_i\left[1+\alpha_i \sin(2\pi ft)+\frac{1}{2}\alpha_i^2 \sin^2(2\pi ft) \right],\quad
\Gamma_o(t) = \Gamma_o\left[1-\alpha_o \sin(2\pi ft)+\frac{1}{2}\alpha_o^2 \sin^2(2\pi ft) \right].
\end{equation}
We note that the average values of the rates over a full driving period are
\begin{equation}
\overline \Gamma_i  = \frac{1}{T}\int_0^Tdt \Gamma_i(t) = \Gamma_i(1+\alpha_i^2/4),\quad
\overline \Gamma_o  =\frac{1}{T}\int_0^Tdt  \Gamma_o(t)  = \Gamma_o(1+\alpha_o^2/4).
\end{equation}
In principle, second-order perturbation theory allows us to derive an analytic expression for the waiting time distribution for all frequencies. However, given the lengthy expressions that the general derivation yields, we focus here on the high-frequency limit, $f \gg \Gamma_i,\Gamma_o$, where all expressions may be simplified considerably. Below we expand each quantity around its corresponding steady-state equivalent with the averaged rates $\overline \Gamma_i$ and $\overline \Gamma_o$ inserted. We neglect terms that are negligible in the high-frequency limit, $f\gg \Gamma_i,\Gamma_o$ and then find
\begin{equation}
p_1(t_0) =  p_1^s(t_0,\overline \Gamma_i,\overline \Gamma_o)\left(1-\frac{\Gamma_o}{2\pi f}\cos\left(2\pi f t_0 \right)(\alpha_i+\alpha_o)\right),
\label{p0 high}
\end{equation}
as well as
\begin{equation}
\label{U11 high}
[\mathbf{U}_{0}(t,t_0)]_{11} = [\mathbf{U}_{0}(t,t_0,\overline \Gamma_i)]_{11}^s\left(1+ \gamma_i+\gamma_i^2/2 \right),\quad
[\mathbf{U}_{0}(t,t_0)]_{22} =  [\mathbf{U}_{0}(t,t_0,\overline \Gamma_o)]_{22}^s\left(1+ \gamma_o+\gamma_o^2/2 \right),
\end{equation}
and
\begin{equation}
\begin{split}
[\mathbf{U}_{0}(t,t_0)]_{21} =& [\mathbf{U}_{0}(t,t_0,\overline \Gamma_i,\overline \Gamma_o)]^s_{21} \Bigg(1+\frac{\cos\left[2\pi f t \right]}{2\pi f}\left[\frac{e^{\Gamma_ot+\Gamma_i t_0}\alpha_i (\Gamma_i-\Gamma_o)}{e^{\Gamma_ot+\Gamma_it_0}-e^{\Gamma_it+\Gamma_ot_0}}-\alpha_o \Gamma_o \right]+\frac{1}{2 (2\pi f)^2}\Bigg[\alpha_i\alpha_o \Gamma_i\Gamma_o\cos\left(2\pi f(t-t_0)\right) \\
&-\frac{2 \alpha_i(\Gamma_i-\Gamma_o)\big(e^{\Gamma_o t+\Gamma_i t_0}\cos\left(2\pi f t \right)-e^{\Gamma_i t+\Gamma_o t_0}\cos\left(2\pi f t_0 \right)\big)\left( \alpha_o\Gamma_o \cos\left(2\pi f t\right)+\alpha_i\Gamma_i \cos\left( 2\pi f t_0  \right) \right)}{e^{\Gamma_ot+\Gamma_it_0}-e^{\Gamma_i t+\Gamma_o t_0}} \Bigg]\Bigg),
\end{split}
\label{U21 high}
\end{equation}
with $\gamma_x \equiv \frac{\alpha_x \Gamma_x }{2\pi f} (\cos[2\pi f t ]-\cos[2\pi f t_0 ])$, $x=i, o$. In the expressions above for the matrix elements of $\mathbf{U}_{0}(t,t_0)$, we have omitted terms that eventually vanish when averaged over a period of the drive. By averaging the idle-time probability in Eq.~\eqref{Pi calc} over a period of the drive, we find
\begin{equation}
\Pi(t) = \Pi_s(t,\overline \Gamma_i,\overline \Gamma_o) \left(1-\frac{1}{2}\left(\frac{1}{2\pi f}\right)^2\frac{\Gamma_i^2 \Gamma_o^2 \left(e^{\Gamma_i t}-e^{\Gamma_o t}\right)}{e^{\Gamma_i t}\Gamma_i^2-e^{\Gamma_o t}\Gamma_o^2}\cos\left(2\pi f t \right)\alpha_o^2 \right),
\end{equation}
and we then arrive at the high-frequency expression for the waiting time distribution in Eq.~(3) of the main text,
\begin{equation}
\mathcal{W}(\tau) = \mathcal{W}_s(\tau,\overline \Gamma_i,\overline\Gamma_o)\left[1+\frac{\alpha_o^2}{2}\cos\left(2\pi f \tau\right) \right].
\end{equation}